\documentclass{emulateapj}
\def\degpoint{\ifmmode ^{\rm{o}}\!. \else $^{\rm{o}}\!.$\fi}

\newcommand{\Msun}{\mbox{M$_{\odot}$}}

\newcommand{\Mjup}{\mbox{M$_{\rm Jup}$}}

\newcommand{\ltsimeq}{\raisebox{-0.6ex}{$\,\stackrel
         {\raisebox{-.2ex}{$\textstyle <$}}{\sim}\,$}}
\newcommand{\gtsimeq}{\raisebox{-0.6ex}{$\,\stackrel
         {\raisebox{-.2ex}{$\textstyle >$}}{\sim}\,$}}

\begin{document}
\title{Detection Limits from the McDonald Observatory Planet Search 
Program}
\author{Robert A.~Wittenmyer, Michael Endl, William D.~Cochran}
\affil{McDonald Observatory, University of Texas at Austin, Austin, TX 
78712}
\email{
robw@astro.as.utexas.edu, mike@astro.as.utexas.edu,
wdc@shiraz.as.utexas.edu}
\author{Artie P.~Hatzes}
\affil{Th\"uringer Landessternwarte Tautenburg, Sternwarte 5, 07778 
Tautenburg, Germany}
\author{G.~A.~H.~Walker, S.~L.~S.~Yang}
\affil{Physics \& Astronomy, University of Victoria, BC  V8W 3P6, Canada}
\author{Diane B.~Paulson}
\affil{NASA/Goddard Space Flight Center, Code 693, Planetary Systems 
Branch, Greenbelt, MD 20771}

\shorttitle{McDonald Observatory Planet Search}
\shortauthors{Wittenmyer et al.}
\slugcomment{Accepted for publication in AJ}
\begin{abstract}

\noindent Based on the long-term radial-velocity surveys carried out with
the McDonald Observatory 2.7m Harlan J.~Smith Telescope from 1988 to the
present, we derive upper limits to long-period giant planet companions for
31 nearby stars.  Data from three phases of the McDonald Observatory 2.7m
planet-search program have been merged together and for 17 objects, data
from the pioneering Canada-France-Hawaii Telescope (CFHT) radial-velocity
program have also been included in the companion-limits determination.  
For those 17 objects, the baseline of observations is in excess of 23
years, enabling the detection or exclusion of giant planets in orbits
beyond 8 AU.  We also consider the possibility of eccentric orbits in our
computations.  At an orbital separation of 5.2 AU, we can exclude on
average planets of M sin $i$ \gtsimeq 2.0$\pm$1.1 \Mjup\ ($e=0$) and M sin
$i$ \gtsimeq 4.0$\pm$2.8 \Mjup\ ($e=0.6$) for 25 of the 31 stars in this
survey.  However, we are not yet able to rule out ``true Jupiters,''
i.e.~planets of M sin $i \sim$ 1 \Mjup\ in 5.2 AU orbits.  These limits
are of interest for the Space Interferometry Mission (SIM), Terrestrial
Planet Finder (TPF), and Darwin missions which will search for terrestrial
planets orbiting nearby stars, many of which are included in this work.

\end{abstract}

\section{Introduction}

In the ten years since the discovery of the first extrasolar planet
orbiting a main-sequence star \citep{mayor95}, more than 170 planets have
been discovered.  The vast majority of these have been detected via
high-precision radial-velocity measurements of the planet's gravitational
influence on its host star.  Current Doppler surveys now routinely achieve
precisions of 2-3 m s$^{-1}$ (e.g.~Cochran et al.~2004), facilitating the
detection of ever-lower-mass companions, such as the ``hot Neptunes''
\citep{bonfils05, mcarthur04, santos04b, butler04} and even
``super-Earths'' \citep{rivera05}.
 
However, there exists a relatively long record of radial velocity data at
somewhat lower precision ($\sim$ 15-20 m s$^{-1}$) which ought not to be
ignored.  Such data now cover nearly a quarter-century \citep{cw79,
campbell88, walker95}, and as such are extremely useful in probing nearby
stars for long-period giant planets akin to our own Jupiter (orbital
period 11.9 years).  These data sets are valuable tools in the search
for extrasolar analogs to our Solar System. We are now beginning to obtain
meaningful answers to some of the following questions: What is the
frequency of long-period giant planets in the solar neighborhood?  What
implications would the lack of such planets have on theories of planet
formation?  How many planetary systems resemble our own Solar system?
 
In this work, we merge radial-velocity data from three phases of
observations at McDonald Observatory, and we also include data from the
CFHT study of \citet{walker95}.  In \S 2, we describe the data sets, \S 3
describes the merging procedure and outlines the algorithm used to derive
companion limits, and in \S 4 we present and discuss the constraints these
data place on the presence of planets orbiting these stars.

\section{Observations}

\subsection{A Brief History of Radial Velocity Planet Detection}

Radial-velocity searches for planets around other stars began well before
the 1995 discovery of 51~Peg\,b. \citet{griffin73} showed that the
fundamental limit to the precision of classical radial velocity techniques
was not instruments used, but rather was a result of the calibration
processes.  They pointed out that the optical paths followed by the
stellar beam and the separate comparison source illuminated the
spectrograph optics differently. \citet{tull69} demonstrated how zonal
errors in spectrograph optics contribute directly to wavelength (and thus
velocity) errors. In addition, thermal and mechanical motion of the
spectrograph in the time interval between obtaining the stellar and
calibration observations also contributed to the measured velocity errors.
\citet{griffin73} proposed that these sources of errors could be defeated
through the calibration of the spectrograph by an absorption line spectrum
imposed on the stellar light {\itshape before} it entered the
spectrograph.  They suggested that the use of the very convenient
absorptions by the Earth's atmospheric O$_2$ band at 6300{\AA} would
permit stellar radial velocities to be measured to a precision perhaps as
good as 0.01\,km\,s$^{-1}$. \citet{cw79} first used an externally applied
absorption spectrum to provide a stable velocity metric for precise radial
velocity measurement as applied to planet detection.  Instead of the
telluric absorption spectrum, they chose to use a stabilized HF gas
absorption cell in front of the coud\'e spectrograph of the Canada France
Hawaii Telescope (CFHT). The HF 3-0 R branch lines near 8700{\AA} gave
radial velocity precision of 15\,m\,s$^{-1}$ in their twelve-year survey
of 21 bright solar-type stars \citep{campbell88,walker95}.  Although
\citet{walker95} did not claim any planet detections based on the CFHT
data alone, two of their target objects ($\epsilon$~Eridani and
$\gamma$~Cephei) were subsequently found to have planetary companions when
the CFHT data were combined with McDonald Observatory data
\citep{hatzes00,hatzes03}.
 
The CFHT survey marked the start of a large number of diverse efforts to
conduct high-precision radial velocity surveys.  At the University of
Arizona, \citet{mcmillan85,mcmillan90,mcmillan94} developed a precise
radial velocity spectrometer based on the use of a Fabry-Perot etalon in
transmission in front of a cross-dispersed echelle spectrograph. This
instrument was able to achieve velocity precision of 8\,m\,s$^{-1}$ for
bright sources \citep{mcmillan94}, and led to the discovery of the radial
velocity variability of K-giants \citep{smith87,cochran88}. The idea of
using various types of interferometers to obtain precise stellar velocity
measurements has been quite popular, with designs proposed by
\citet{cochran82}, \citet{connes85}, \citet{angel97}, and \citet{ge02}. Of
these designs, only \citet{vaneyken03} were successful in developing the
design into an instrument that actually discovered an extrasolar planet 
\citep{ge06}.
 
To our knowledge, \citet{beckers76} first used the molecular iodine
B$^3\Pi_{0u}^+$--X$^1\Sigma_g^+$ band as a velocity reference system, in
application to his measurements of the solar rotation. \citet{koch84} then
extended this use of the B-X I$_2$ band to the measurement of the Doppler
shifts of several solar photospheric absorption lines. \citet{li88} then
extended the use of I$_2$ as a velocity metric in searching for p-mode
oscillations of $\alpha$~CMi. \citet{marcy92} then implemented the use of
the I$_2$ velocity metric for large-scale precise radial velocity surveys
for extrasolar planets.  The use of an I$_2$ cell  offers the major 
advantage that the I$_2$ lines can be used to measure the spectrometer 
instrumental profile \citep{valenti95} and thus measure very precise 
radial velocity variations \citep{butler96}.  An alternative is to use an 
optical-fiber fed cross dispersed echelle spectrograph with simultaneous 
wavelength calibration.  This was first done with ELODIE \citep{baranne96} 
and later with ELODIE \citep{queloz00}.  To further improve precision with 
this technique, the HARPS instrument \citep{rupp04} is placed in a 
temperature-stabilized vacuum chamber.
 
\subsection{The McDonald Observatory Planetary Search Program}

The McDonald Observatory Planetary Search program comprises a large,
multifaceted investigation to detect and characterize planetary companions
to other stars in our galaxy.  It began in 1988 as a high-precision radial
velocity survey of bright nearby stars using the McDonald Observatory 2.7m
Harlan J.~Smith Telescope and coud\'e spectrograph, but has expanded
substantially in size and scope since then. Phase I of the radial velocity
planet search program used the telluric O$_2$ lines near 6300{\AA} as the
velocity metric, a technique suggested by \citet{griffin73}.  A single
order of the McDonald 2.7m telescope coud\'e echelle spectrograph (cs12)
was isolated onto a Texas Instruments $800 \times 800$ CCD at $R =
210,000$.  This system gave about 20\,m\,s$^{-1}$ precision on stars down
to $V = 6$, but suffered from systematic velocity errors, possibly due to
prevailing atmospheric winds.  Diurnal and seasonal variability in the
winds introduced spurious periodic signals in the data.  We therefore
switched to a temperature stabilized I$_2$ cell as the velocity metric in
1992.  This eliminated the systematic errors, and gave a routine radial
velocity precision of $\sim$15\,m\,s$^{-1}$.  This precision was limited
by the 9.6{\AA} bandpass of the single order of the echelle grating, and
by the poor charge-transfer and readout properties of the TI 800x800 CCD.
To solve these problems, and to achieve substantially improved precision,
we began Phase III of the radial velocity program in July 1998, using the
same I$_2$ cell with the newly installed 2dcoud\'e cross-dispersed echelle
spectrograph \citep{tull94} with its 2048x2048 Tektronix CCD.  We set up
the spectrograph to include echelle orders from 3594{\AA} to 10762{\AA}, 
which covers the Ca~II H and K lines used to measure stellar activity.
Wavelength coverage is complete from the blue end to 5691{\AA}, and there
are increasingly large inter-order gaps from there to the red end
\citep{tull95}.  Using the full 1200{\AA} bandpass of the I$_2$ absorption
band at the $R = 60,000$ focus of the 2dcoud\'e allows routine 
internal precision
of 6--9\,m\,s$^{-1}$ to be achieved.  The McDonald program has
subsequently been expanded significantly to include a collaboration with
Martin K\"{u}rster and colleagues on southern hemisphere radial velocity
surveys using the facilities of ESO at La Silla and the VLT
\citep{martin00, endl02, martin2003}, and a program to use the Keck~1
HIRES spectrograph to search for planetary companions to stars in the
Hyades \citep{cochran02, paulson02, paulson03, paulson04a, paulson04b} and
in the use of the High Resolution Spectrograph on the Hobby-Eberly
Telescope \citep{cochran04}.  Observations from the McDonald planet search
program have been used in the discovery of four planets: 16 Cyg Bb
\citep{cochran97}, $\gamma$ Cep Ab \citep{hatzes03}, $\epsilon$ Eridani b
\citep{hatzes00}, and HD~13189b \citep{hatzes05}.  McDonald data were also
instrumental in the detection of the brown dwarf HD~137510b
\citep{endl04}.
 
\subsection{Observational Data Presented Here}

The majority of the data used in this study were obtained by the McDonald
Observatory program Phases I-III using the 2.7m Harlan J.~Smith telescope.  
A list of the target stars is given in Table 1. For 17 of the 31 stars in
this study, additional data were available from the CFHT precision
radial-velocity work of \citet{walker95}.  Table 2 gives a summary of all
four data sets, including the total time span, rms of each data set, and
the chromospheric emission ratio log $R'_{HK}$ \citep{noyes84} computed
from the McDonald Phase III measurements of the Ca~II S-index.  The rms 
values listed in Table 2 include the internal uncertainties of 
6--9\,m\,s$^{-1}$ and the velocity jitter inherent to each star.

\section{Data Analysis}

\subsection{Merging the Data}

Since each of the four data sets consists of velocities measured relative
to an independent and arbitrary zero-point, it was necessary to implement
a consistent and robust method of combining them. Before merging, an
iterative outlier-rejection routine was applied to each data set
separately.  Data points which were more than $3.3 \times$RMS from the
mean were rejected.  This criterion corresponds to a 99.9\% confidence
level for a Gaussian distribution.  The mean and RMS were then re-computed
and the outlier rejection was repeated until no more points were rejected.  
As the targets are well-studied constant-velocity stars, we are confident
that we have a well-defined distribution about the mean.  For the visual
binary 70~Oph~A, several velocities were systematically too low due to
spectral contamination caused by the chance alignment of both components
on the slit, and these points were also removed.  The fainter component
($\Delta m = 1.8$ mag)  adds a Doppler-shifted second spectrum, which
distorts the line shapes.  Since our velocity computation method assumes a
single set of lines, the resultant velocity is skewed by this
contamination.

The merging of these data sets was accomplished in the following manner:  
McDonald Phases II and III are joined with a trial offset, and a
least-squares linear fit of a trend of velocity with time is performed on
the combination.  The offset between the Phase II and III data sets which
minimizes the RMS about that linear fit is the one which is applied in
order to merge two phases together.  This process was then repeated
sequentially to join the Phase I and then the CFHT data to the growing
data string.  The least-squares fitting allows for a linear trend to be
present; the merging process was re-done for all non-binaries, this time
forcing the slope to be zero (i.e. minimizing the RMS about the mean of
the combined data for a grid of trial offsets).  The RMS about the mean of
data merged in this manner was compared to the RMS about a linear fit to
data merged allowing a trend.  Since our null hypothesis is that the stars
are radial-velocity constants, the method which allowed a slope was only
chosen for the 9 stars which showed an improvement in the RMS of the
merged data set when a slope was allowed.  Those stars were: $\eta$ Cep,
$\eta$ Cas, $\alpha$ For, $\theta$ UMa, $\xi$ Boo~A, $\mu$ Her, 70 Oph A,
and 61~Cyg~A \& B.  All of these stars are well-known long-period
binaries, so the use of a slope was justified.

The orbital periods of these 9 known binaries were sufficiently long that
a simple linear approximation was sufficient, except for $\mu$ Her, $\xi$
Boo A (quadratic trend) and 70~Oph~A (Keplerian solution).  Fits to these
trends were subtracted from the data sets and the residuals were used as
input for the detection-limit algorithm.  The three data sets on the
visual binary $\mu$ Her were merged in a similar manner.  The binary
period (43.2 years; Worley \& Heintz 1995) is such that a linear trend is
a suboptimal approximation for the purposes of merging the 16.9 years of
data, and a quadratic fit was used to approximate the shape of the orbit.  
This procedure was also used for $\xi$ Boo~A; the data and fits are shown
in Figure 1.  For 70~Oph~A, the quadratic approximation was still
insufficient, as the 16.1 years of available data represents a significant
portion of the 91.44 year published orbital period \citep{batten84}, and
an orbital solution for the binary system had to be obtained using
GaussFit \citep{jefferys87}.  The solution to the combined data set
(McDonald Phases I-III plus much lower-precision data from Batten et
al.~1984) is given in Table 3, and plotted in Figure 2.  The large
uncertainties in the fitted parameters are due to the fact that the
available data do not encompass a complete orbit, and hence the model is
poorly constrained.  The fitted value of $\omega$ was derived by a
rudimentary grid search around the published value, as GaussFit would not
converge with $\omega$ as a free parameter.  The orbital parameters of the
binary system are substantially different than those reported in
\citet{batten84}, likely owing to the fact that the McDonald velocities,
which are an order of magnitude more precise, drive the fit.  Only the
McDonald data were used in the computation of companion limits for 70 
Oph~A.

\subsection{Periodogram Analysis}

After the data sets were merged together, and trends due to binary orbits
were subtracted, we searched for periodicities using a Lomb-Scargle
periodogram \citep{lomb76, scargle82}.  False alarm probabilities (FAP)
were established using a bootstrap randomization method as described in
\citet{kurster97} and \citet{endl02}.  This bootstrap method, unlike the
nominal analytic FAP formula of \citet{scargle82}, makes no assumptions
about the distribution of the data; rather, the error distribution of the
actual data themselves is used.  The periodogram search interval was from
2 days up to the full extent of observations for each object.  We used
10,000 bootstrap randomizations to determine the FAP of the highest peak
in the periodogram for each star; the results are given in Table 4.  Only
one star had a periodicity with a significance greater than 99\%: $\pi^3$
Ori, with a peak at 73.26 days and a bootstrap FAP of 0.8\%.  As this is
an active early-type (F6V) star, we are confident that this is due to a
combination of stellar activity and reduced velocity precision (from the
paucity of lines) rather than the presence of a companion.  Additionally,
no peak was evident in a periodogram on the more precise McDonald Phase
III data alone, supporting our conclusion that it is a spurious noise
spike.

\subsection{Determination of Companion Limits}

Companion limits were determined via an algorithm which injected test signals
into the data, and then attempted to recover that signal using a periodogram
search.  Test signals were generated using the method of Lehmann-Filh\'es 
\citep{lehmann} as described in \citet{binnendijk}:

\begin{equation}
V_r=\gamma + K[e~cos~\omega + cos(\nu + \omega)],
\end{equation}

\noindent where $V_r$ is the radial velocity, $\gamma$ is the systemic
velocity, $K$ is the velocity semiamplitude, $e$ the orbital eccentricity,
$\omega$ the argument of periastron, and $\nu$ the true anomaly. In the
above equation, $\nu$ can be expressed in terms of observables via the
following relations:

\begin{equation}
tan~\frac{\nu}{2}=\Big[\frac{(1+e)}{(1-e)}\Big]^{1/2}~tan~\frac{E}{2}
\end{equation}

\noindent and

\begin{equation}
M=\frac{2\pi}{P}(t-T_0)=E-e~sin~E,
\end{equation}

\noindent for a signal with period $P$, periastron passage time $T_0$,
observation time $t$, mean anomaly $M$, and eccentric anomaly $E$. For
each data set, the algorithm stepped through 300 trial periods at even
steps in the logarithm between 2 days and the total duration of
observations.  At each trial period, the program sampled 30 phase steps of
the periastron time $T_0$, and for nonzero eccentricities, sampled 36
values of the argument of periastron $\omega$ at intervals of 10 degrees.  
The velocity semi-amplitude $K$ of the test signal was allowed to vary
from 10 to 1000 m s$^{-1}$.  The systemic velocity $\gamma$ of each
combined data set was forced to be equal to zero.  For each set of
Keplerian orbital parameters, synthetic radial velocities were generated
using the observation times from the input data.  This simulated signal
was then added to the data, and a Lomb-Scargle periodogram was used to
attempt to recover that signal.  For a signal to count as having been
recovered, the periodogram's highest peak had to occur at the correct
period with a FAP of less than 0.1\%.  The FAP was estimated using the
formula from \citet{hb86}.  For each $K$ velocity, 99\% of the 1080 test
signals\footnote{For signals with $e=0$, \textit{all} 30 variations in
$T_0$ had to be recovered, as $\omega$ had no meaning.} had to be
recovered in this manner in order for that velocity semiamplitude
(corresponding to a planet of a given mass at a given semimajor axis) to
be considered ``ruled out'' by the data.  If more than 1\% of the orbital
configurations tested at a given $K$ value were \textit{not} ruled out in
this manner, the algorithm increased $K$ by 1 m s$^{-1}$ and repeated the
process until it was able to recover 99\% of the parameter configurations
($e, \omega, T_0$) that were tested.  Once this occurred, the planetary
mass ruled out was computed by the following:

\begin{equation}
M_2~sin~i = (1-e^2)^{1/2}\bigg[(1.036\times 
10^{-7})M_1^{2}PK^{3}\bigg]^{1/3},
\end{equation}

\noindent where $M_1$ is the stellar mass in solar masses, $P$ is the
orbital period in days, $K$ is the radial-velocity semiamplitude in km
s$^{-1}$, and $M_2$ sin $i$ is the projected planetary mass in solar
masses.  The algorithm then moved to the next trial period and repeated
the entire process.

Unlike most previous companion limit determinations
\citep{murdoch93,walker95, nelson98, cumming99, endl02}, this procedure
allows for nonzero eccentricities in the trial orbits (but see Desidera et
al.~2003).  For the eccentric case, a value of $e=0.6$ was chosen; of the 
known extrasolar planets, 90\% have
$e<0.6$ \citet{marcy05}.  Allowing higher eccentricities substantially
reduced the ability to rule out the test signals, as the sporadic sampling
of the data would likely miss the relatively high-velocity points near
periastron.  Figure 3 shows the effect of allowing various ranges of
eccentricity.  Note that although the limits derived by this study are
rendered somewhat less stringent by the inclusion of nonzero
eccentricities, the effect is relatively minor for our adopted value of
$e=0.6$.  Allowing a larger range of eccentricities (up to $e=0.9$),
however, substantially reduces the sensitivity of the companion-limit
determination, as demonstrated in Figure 3.  We emphasize that the upper
limits derived by the method described above are much more stringent, and
hence will result in higher companion-mass limits than those reported by 
previous studies.

\section{Results and Discussion}

Limits to planetary companions derived using these data are shown in
Figures 4-11.  In each panel, the lower set of points (solid line)  
represents the companion limits for the zero-eccentricity case, and the
dotted line is for the case of $e=0.6$. Notably, despite the abundance and
quality of data available in this study, we are as yet unable to rule out
\textit{any} planets with M sin $i \ltsimeq$ 1 \Mjup\ in 5.2 AU orbits
with eccentricities as large as $e=0.6$.  When only circular orbits are
considered, such objects can be ruled out for $\tau$ Ceti, $\sigma$ Dra,
61 Cyg A, and 61 Cyg B.  Table 5 lists the minimum planet masses that can
be excluded by these data at selected semimajor axes, for the $e=0$ and
$e=0.6$ cases.  The results given in Table 5 are shown in histogram format
in Figure 12, which indicates that for most stars in this survey,
Saturn-mass planets in close orbits ($a \sim 0.1$ AU) can be ruled out.

It is also useful to consider the effect of giant planets in intermediate
orbits ($a \sim 2-3$ AU) which may perturb lower-mass planets within the
habitable zone of the star.  If such objects can be excluded with
confidence, their host stars become attractive candidates for the
\textit{Terrestrial Planet Finder} (TPF) and Darwin missions, which aim to
detect Earth-like planets in the habitable zone.  \citet{menou03} defined
a planet's zone of influence to extend from $R_{in}=(1-e)a-3R_{Hill}$ to
$R_{out}=(1+e)a+3R_{Hill}$, where the Hill radius is

\begin{equation}
R_{Hill}=a\Bigg(\frac{M_p}{3M_*}\Bigg)^{1/3},
\end{equation}

\noindent and $e$ is the planet's eccentricity, $a$ is its semimajor axis,
$M_p$ is its mass, and $M_*$ is the mass of the star.  Simulations by
\citet{menou03} demonstrated that terrestrial planets were nearly always
ejected or consumed in systems where an eccentric giant planet's zone of
influence overlapped the habitable zone.  We can then ask whether the
companion limits derived in this work can be used to exclude such
perturbing bodies.  Such a pursuit is limited by the fact that even
distant giant planets can disrupt the habitable zone if their orbits are
sufficiently eccentric, and as shown in Figure 1, the nature of the
radial-velocity data is such that we are least sensitive to the most
eccentric planets.  Nevertheless, for small eccentricities ($e\ltsimeq
0.2$), it is possible to combine these dynamical calculations with our
companion-limit determinations to define a ``safe zone'': a region of
parameter space in which we can exclude perturbing giant planets exterior
to the habitable zone.  For such regions, the possibility of terrestrial
planets in the habitable zone remains open for programs such as TPF and
Darwin.  In Figures 4-11, the region to the left of the dot-dashed line
and above the dotted line defines the ``safe zone'' for perturbing outer
giant planets with $e<0.2$.  These were only plotted for main-sequence
stars, using the definition of the ``continuously habitable zone'' given
in \citet{kasting93}.  The region left of the dot-dashed line and
\textit{below} our limits represents a set of potentially dangerous
objects which would disrupt the habitable zone, yet be undetectable with
the current data.  Higher eccentricities would push the dot-dashed line to
the right and reduce its slope, such that for perturbers with $e\gtsimeq
0.5$, our limits computations can say nothing about such objects (i.e.~the
curves would not intersect).  The companion limits we have derived thus
place some constraints on potentially disruptive objects in these systems,
which will assist in target selection for the TPF and Darwin missions.

Noting that the Phase III data are of substantially higher quality than
the previous data sets, we asked what velocity rms would be required to
rule out a Jupiter analog orbiting a solar-type star.  We generated
simulated observations consisting of Gaussian noise at the actual
observation times (spanning 16 years) for 16~Cyg~A, the star in this study
which is closest in spectral type (G1.5 V) to our Sun.  Figure 13 shows
the results of the companion-limit algorithm on four of these simulated
data sets with four levels of rms scatter.  In order to rule out a planet
with M sin $i$ of 1 \Mjup\ in a 5.2 AU orbit ($e=0.1$), the data need to
have an rms less than about 10 m s$^{-1}$.  The rms of Phase III
observations of 16~Cyg~A is 6.9 m s$^{-1}$ over a period of 6.3 years,
whereas the complete 16 years has an overall rms of 26.8 m s$^{-1}$.  
Hence, a Jupiter analog could be ruled out for 16~Cyg~A with about 10 more
years of data of the same quality as McDonald Phase III.  The weighted
mean rms of all McDonald Phase III observations in this survey is 12.6 m
s$^{-1}$.  Eleven stars (see Table 2) currently achieve a Phase III rms
better than 8 m s$^{-1}$; this represents one-third of the stars discussed
in this work.  These simulations show that the average precision of Phase
III needs to be improved by about 2-3 m s$^{-1}$ in order to achieve the
sensitivity required to detect or exclude Jupiter analogs for all of these
stars.

\section{Summary}

We have amassed a substantial quantity of radial-velocity data on 31 nearby
stars, spanning up to 23 years of observations.  We have applied a robust
method to place conservative limits on planetary companions orbiting these
stars.  We have considered the effect of eccentric orbits in our computations,
which better reflects the diversity of known extrasolar planetary systems.  
Saturn-mass planets within 0.1~AU can be ruled out for nearly all of these
stars.  At 5.2 AU, we can exclude on average planets with M sin $i$ \gtsimeq
2.0$\pm$1.1 \Mjup\ ($e=0$)and M sin $i$ \gtsimeq 4.0$\pm$2.8 \Mjup\ ($e=0.6$)  
for 25 of the 31 stars.  Although we now have a quite sufficient time baseline
for the detection of Jupiter analogs in $\sim$12 year orbits, we find that the
overall velocity precision is not yet sufficient to exclude Jupiter-mass
planets at 5.2~AU.  However, improvements in the current McDonald Observatory
2.7m telescope planet search program put the desired level of precision within
reach for inactive stars.

\acknowledgements

This research was supported by NASA grants NNG04G141G and NNG05G107G.  
R.W.~acknowledges support from the Sigma Xi Grant-in-Aid of Research and
the Texas Space Grant Consortium.  D.B.P.~is currently a National Research
Council fellow working at NASA's Goddard Space Flight Center.  We are
grateful to Barbara McArthur for her assistance with GaussFit software,
and to E.L.~Robinson for insightful comments which improved this
manuscript.  This research has made use of NASA's Astrophysics Data System
(ADS), and the SIMBAD database, operated at CDS, Strasbourg, France, as
well as computing facilities at San Diego State University.



\begin{deluxetable}{llllll} 
\tabletypesize{\scriptsize}
\tablecolumns{6}
\tablewidth{0pt}
\tablecaption{Target List and Stellar Parameters \label{tbl-1}}
\tablehead{
\colhead{Star} & \colhead{HR} & \colhead{Spec.~Type} & \colhead{$V$ 
magnitude} & \colhead{Mass (\Msun)} & \colhead{Reference for Mass 
Estimate}}

\startdata
$\eta$ Cas & 219 & G0V & 3.44 & 0.90 & \citet{carlos99} \\
$\tau$ Cet & 509 & G8V & 3.50 & 0.65 & \citet{santos04a} \\
$\theta$ Per & 799 & F8V & 4.12 & 1.15 & \citet{carlos99} \\
$\iota$ Per & 937 & G0V & 4.04 & 1.23 & \citet{carlos99} \\
$\alpha$ For & 963 & F8V & 3.87 & 1.30 & \citet{carlos99} \\
$\kappa^1$ Cet & 996 & G5V & 4.82 & 1.05 & \citet{carlos99} \\
$\delta$ Eri & 1136 & K0IV & 3.54 & 0.96 & \citet{carlos99} \\
$o^2$ Eri & 1325 & K1V & 4.43 & 0.65 & \citet{santos04a}\\
$\pi^3$ Ori & 1543 & F6V & 3.19 & 1.24 & \citet{carlos99} \\
$\lambda$ Aur & 1729 & G1.5IV-V & 4.71 & 1.15 & \citet{carlos99} \\
$\theta$ UMa & 3775 & F6IV & 3.17 & 1.53 & \citet{carlos99} \\
36 UMa & 4112 & F8V & 4.82 & 1.14 & \citet{carlos99} \\
$\beta$ Vir & 4540 & F8V & 3.61 & 1.36 & \citet{carlos99} \\
$\beta$ Com & 4983 & G0V & 4.28 & 1.05 & \citet{carlos99} \\
61 Vir & 5019 & G6V & 4.75 & 1.01 & \citet{carlos99} \\
$\xi$ Boo A & 5544 & G8V & 4.55 & 0.86 & \citet{f98} \\
$\lambda$ Ser & 5868 & G0V & 4.43 & 0.97 & \citet{carlos99} \\
$\gamma$ Ser & 5933 & F6V & 3.85 & 1.28 & \citet{carlos99} \\
36 Oph A & 6402 & K1V & 5.29 & 0.78 & \citet{walker95} \\
$\mu$ Her & 6623 & G5IV & 3.42 & 1.10 & \citet{subgiants}\\
70 Oph A & 6752 & K0V & 4.03 & 0.97 & \citet{carlos99} \\
$\sigma$ Dra & 7462 & K0V & 4.68 & 0.85 & \citet{nelson98} \\
16 Cyg A & 7503 & G1.5V & 5.96 & 0.98 & \citet{carlos99} \\
31 Aql & 7373 & G8IV & 5.16 & 0.95 & \citet{subgiants}\\
$\beta$ Aql & 7602 & G8IV & 3.71 & 1.50 & \citet{subgiants}\\
$\gamma^2$ Del & 7948 & K1IV & 4.27 & 1.90 & \citet{subgiants}\\
$\eta$ Cep & 7957 & K0IV & 3.43 & 1.39 & \citet{carlos99} \\
61 Cyg A & 8085 & K5V & 5.21 & 0.67 & \citet{walker95} \\
61 Cyg B & 8086 & K7V & 6.03 & 0.59 & \citet{walker95} \\
HR 8832 & 8832 & K3V & 5.56 & 0.79 & \citet{nelson98} \\
$\iota$ Psc & 8969 & F7V & 4.13 & 1.38 & \citet{carlos99} \\

\enddata
\end{deluxetable}

\begin{deluxetable}{llllllll}
\tabletypesize{\scriptsize}
\tablecolumns{8}
\tablewidth{0pt}
\tablecaption{Summary of Observations \label{tbl-2}}
\tablehead{
\colhead{Star} & \colhead{$N$} & \colhead{$T$} & \colhead{CFHT 
rms} & \colhead{Phase I rms} & \colhead{Phase II rms} & \colhead{Phase III 
rms} & \colhead{log $R'_{HK}$}\\
\colhead{} & \colhead{} & \colhead{(years)} & \colhead{(m s$^{-1}$)} & 
\colhead{(m s$^{-1}$)} & \colhead{(m s$^{-1}$)} & \colhead{(m s$^{-1}$)} & 
\colhead{}  }
                                                                                
\startdata

$\eta$ Cas & 131 & 16.3 & \nodata & 28.1 & 28.9 & 7.3 & -4.926\\
$\tau$ Cet & 183 & 23.3 & 14.2 & 22.3 & 28.7 & 10.4 & -4.979\\
$\theta$ Per & 65 & 9.5 & \nodata & \nodata & 59.6 & 17.9 & -4.919\\
$\iota$ Per & 165 & 23.9 & 18.2 & 30.0 & 18.6 & 10.2 & -5.041\\
$\alpha$ For & 66 & 15.3 & \nodata & 28.9 & 41.3 & 15.8 & -5.023\\
$\kappa^1$ Cet & 134 & 23.0 & 23.7 & 34.4 & 29.7 & 20.1 & -4.441\\
$\delta$ Eri & 109 & 16.1 & \nodata & 17.8 & 22.5 & 9.4 & -5.228\\
$o^2$ Eri & 138 & 22.1 & 18.6 & 30.4 & 18.0 & 14.6 & -4.951\\
$\pi^3$ Ori & 160 & 15.9 & \nodata & 169.9 & 113.3 & 26.4 & -4.716\\
$\lambda$ Aur & 63 & 10.4 & \nodata & \nodata & 20.6 & 10.1 & -5.051\\
$\theta$ UMa & 268 & 18.3 & 18.8 & 43.1 & 43.7 & 14.8 & -5.608\\
36 UMa & 194 & 23.0 & 21.0 & 19.4 & 22.4 & 8.2 & -4.811\\
$\beta$ Vir & 205 & 23.3 & 28.4 & 30.5 & 32.6 & 7.6 & -4.942\\
$\beta$ Com & 191 & 22.4 & 18.4 & 31.7 & 42.0 & 11.1 & -4.749\\
61 Vir & 149 & 23.0 & 18.4 & 25.8 & 31.0 & 7.8 & -5.030\\
$\xi$ Boo A & 186 & 23.3 & 23.6 & 34.5 & 31.9 & 23.3 & -4.420\\
$\lambda$ Ser & 63 & 11.5 & \nodata & \nodata & 7.9 & 19.4 & -4.936\\
$\gamma$ Ser & 149 & 15.1 & \nodata & 84.7 & 41.6 & 25.6 & -4.934\\
36 Oph A & 91 & 21.9 & 20.1 & 21.5 & 33.1 & 16.3 & -4.614\\
$\mu$ Her & 173 & 16.9 & \nodata & 28.4 & 24.4 & 9.2 & -5.092\\
70 Oph A & 98 & 16.2 & \nodata & 111.4 & 43.3 & 17.4 & -4.736\\
$\sigma$ Dra & 178 & 23.6 & 14.5 & 21.5 & 23.1 & 9.9 & -4.865\\
16 Cyg A & 102 & 16.0 & \nodata & 34.4 & 29.0 & 6.9 & -5.018\\
31 Aql & 38 & 7.9 & \nodata & \nodata & 22.5 & 11.9 & -5.123\\
$\beta$ Aql & 183 & 22.1 & 14.6 & 28.0 & 20.1 & 16.4 & -5.171\\
$\gamma^2$ Del & 103 & 15.0 & \nodata & 23.7 & 21.7 & 16.8 & -5.354\\
$\eta$ Cep & 187 & 23.5 & 19.2 & 29.5 & 16.4 & 9.6 & -5.223\\
61 Cyg A & 143 & 23.4 & 20.7 & 22.3 & 13.5 & 7.2 & -4.862\\ 
61 Cyg B & 121 & 22.4 & 16.9 & 23.5 & 16.2 & 3.9 & -4.962\\ 
HR 8832 & 119 & 22.9 & 14.9 & 22.8 & 13.8 & 12.9 & -5.013\\
$\iota$ Psc & 54 & 11.2 & \nodata & \nodata & 26.4 & 10.3 & -4.915\\

\enddata
\end{deluxetable}

\begin{deluxetable}{lll}
\tabletypesize{\scriptsize}
\tablecolumns{3}
\tablewidth{0pt}
\tablecaption{Orbital Solution for 70 Oph A \label{tbl-3}}
\tablehead{
\colhead{Parameter} & \colhead{Estimate} & \colhead{Uncertainty} }
\startdata
Period (years) & 179.1 & 99.5 \\
$T_0$ (HJD) & 2445610.22 & 91.74 \\
$e$ & 0.69 & 0.11 \\
$\omega$ (degrees) & 193.6 & \nodata \\
$K_A$ (m s$^{-1}$) & 4163 & 257 \\

\enddata
\end{deluxetable}

\begin{deluxetable}{lrr}
\tabletypesize{\scriptsize}
\tablecolumns{3}
\tablewidth{0pt}
\tablecaption{Results of Periodogram Analysis \label{tbl-4}}
\tablehead{
\colhead{Star} & \colhead{Period (days)} & \colhead{FAP}}
 
\startdata
$\eta$ Cas & 2.964 & 0.296 \\
$\tau$ Cet & 8.389 & 0.034 \\
$\theta$ Per & 348.432 & 0.443 \\
$\iota$ Per & 91.996 & 0.027 \\
$\alpha$ For & 8.262 & 0.395 \\
$\kappa^1$ Cet & 33.069 & 0.034 \\
$\delta$ Eri & 22.051 & 0.788 \\
$o^2$ Eri & 7.463 & 0.900 \\
$\pi^3$ Ori & 73.260 & 0.008 \\
$\lambda$ Aur & 2.549 & 0.188 \\
$\theta$ UMa & 2.587 & 0.178 \\
36 UMa & 3703.704 & 0.694 \\
$\beta$ Vir & 3.307 & 0.262 \\
$\beta$ Com & 9.946 & 0.315 \\
61 Vir & 9.777 & 0.072 \\
$\xi$ Boo A & 6.298 & 0.202 \\
$\lambda$ Ser & 4.929 & 0.773 \\
$\gamma$ Ser & 8.961 & 0.137 \\
36 Oph A & 2.641 & 0.565 \\
$\mu$ Her & 7.623 & 0.039 \\
70 Oph A & 6.138 & 0.989 \\
$\sigma$ Dra & 30.553 & 0.109 \\
16 Cyg A & 2.873 & 0.038 \\
31 Aql & 8.374 & 0.536 \\
$\beta$ Aql & 2.800 & 0.019 \\
$\gamma^2$ Del & 3.837 & 0.099 \\
$\eta$ Cep & 5.940 & 0.030 \\
61 Cyg A & 6.570 & 0.159 \\
61 Cyg B & 6.872 & 0.846 \\
HR 8832 & 25.227 & 0.144 \\
$\iota$ Psc & 2.499 & 0.144 \\
 
\enddata
\end{deluxetable}

\begin{deluxetable}{lrrrrrr}
\tabletypesize{\scriptsize}
\tablecolumns{7}
\tablewidth{0pt}
\tablecaption{Minimum-Mass Companion Limits \label{tbl-5}}
\tablehead{
\colhead{Star} & \colhead{M sin i (\Mjup)} & \colhead{M sin i (\Mjup)} &
\colhead{M sin i (\Mjup)} & \colhead{M sin i (\Mjup)} & \colhead{M sin i 
(\Mjup)} &  \colhead{M sin i (\Mjup)}\\
\colhead{} & \colhead{0.05 AU} & \colhead{0.1 AU} & \colhead{3 AU} & 
\colhead{3 AU} & \colhead{5.2 AU} & \colhead{5.2 AU}\\
\colhead{} & \colhead{$e=0.0$} & \colhead{$e=0.0$} & \colhead{$e=0.6$} &
\colhead{$e=0.0$} & \colhead{$e=0.6$} & \colhead{$e=0.0$}  }
\startdata
$\eta$ Cas & 0.13 & 0.17 & 1.30 & 0.87 & 2.31 & 1.37 \\
$\tau$ Cet & 0.09 & 0.16 & 1.14 & 0.67 & 1.34 & 0.90 \\
$\theta$ Per & 0.48 & 0.62 & 5.71 & 3.46 & \nodata & \nodata \\
$\iota$ Per & 0.12 & 0.16 & 1.46 & 0.95 & 2.70 & 1.69 \\
$\alpha$ For & 0.32 & 0.57 & 6.12 & 3.05 & 8.46 & 4.37 \\
$\kappa^1$ Cet & 0.18 & 0.27 & 2.34 & 1.44 & 3.81 & 2.05 \\
$\delta$ Eri & 0.13 & 0.19 & 2.10 & 1.07 & 12.74 & 1.88 \\
$o^2$ Eri & 0.12 & 0.16 & 1.33 & 0.88 & 1.66 & 1.16 \\
$\pi^3$ Ori & 0.84 & 1.51 & 9.79 & 6.01 & 46.70 & 8.54 \\
$\lambda$ Aur & 0.18 & 0.24 & 2.19 & 1.25 & \nodata & \nodata \\
$\theta$ UMa & 0.24 & 0.36 & 3.09 & 1.96 & 4.60 & 2.57 \\
36 UMa & 0.13 & 0.17 & 1.46 & 0.91 & 2.54 & 1.71 \\
$\beta$ Vir & 0.16 & 0.23 & 2.49 & 1.57 & 4.19 & 2.61 \\
$\beta$ Com & 0.16 & 0.26 & 1.95 & 1.24 & 3.29 & 2.22 \\
61 Vir & 0.14 & 0.20 & 1.62 & 1.16 & 2.58 & 1.69 \\
$\xi$ Boo A & 0.16 & 0.19 & 2.00 & 1.24 & 2.61 & 1.78 \\
$\lambda$ Ser & 0.16 & 0.21 & 2.01 & 1.14 & \nodata & \nodata \\
$\gamma$ Ser & 0.44 & 0.49 & 4.15 & 2.97 & 10.89 & 5.80 \\
36 Oph A & 0.19 & 0.23 & 2.24 & 1.39 & 5.83 & 2.33 \\
$\mu$ Her & 0.14 & 0.21 & 1.53 & 0.96 & 2.55 & 1.52 \\
70 Oph A & 0.46 & 0.84 & 8.71 & 5.04 & 12.78 & 7.19 \\
$\sigma$ Dra & 0.11 & 0.12 & 1.12 & 0.79 & 1.60 & 1.03 \\
16 Cyg A & 0.18 & 0.32 & 2.21 & 1.38 & 5.44 & 2.45 \\
31 Aql & 0.22 & 0.38 & \nodata\tablenotemark{a} & 1.90 & \nodata & \nodata \\
$\beta$ Aql & 0.12 & 0.18 & 1.61 & 1.04 & 2.67 & 1.77 \\
$\gamma^2$ Del & 0.22 & 0.32 & 2.55 & 1.60 & 4.70 & 2.32 \\
$\eta$ Cep & 0.13 & 0.17 & 2.02 & 0.86 & 2.41 & 1.52 \\
61 Cyg A & 0.09 & 0.14 & 1.60 & 0.85 & 2.10 & 0.98 \\
61 Cyg B & 0.07 & 0.10 & 1.12 & 0.66 & 1.48 & 0.80 \\
HR 8832 & 0.11 & 0.14 & 1.65 & 0.81 & 2.56 & 1.14 \\
$\iota$ Psc & 0.26 & 0.38 & 4.35 & 1.85 & 4.91 & 2.64 \\
\enddata
\tablenotetext{a}{Too few data points for a reliable periodogram search, 
due to undersampling of eccentric test signals. }
\end{deluxetable}

\epsscale{1.00}
\clearpage
\begin{figure}
\plottwo{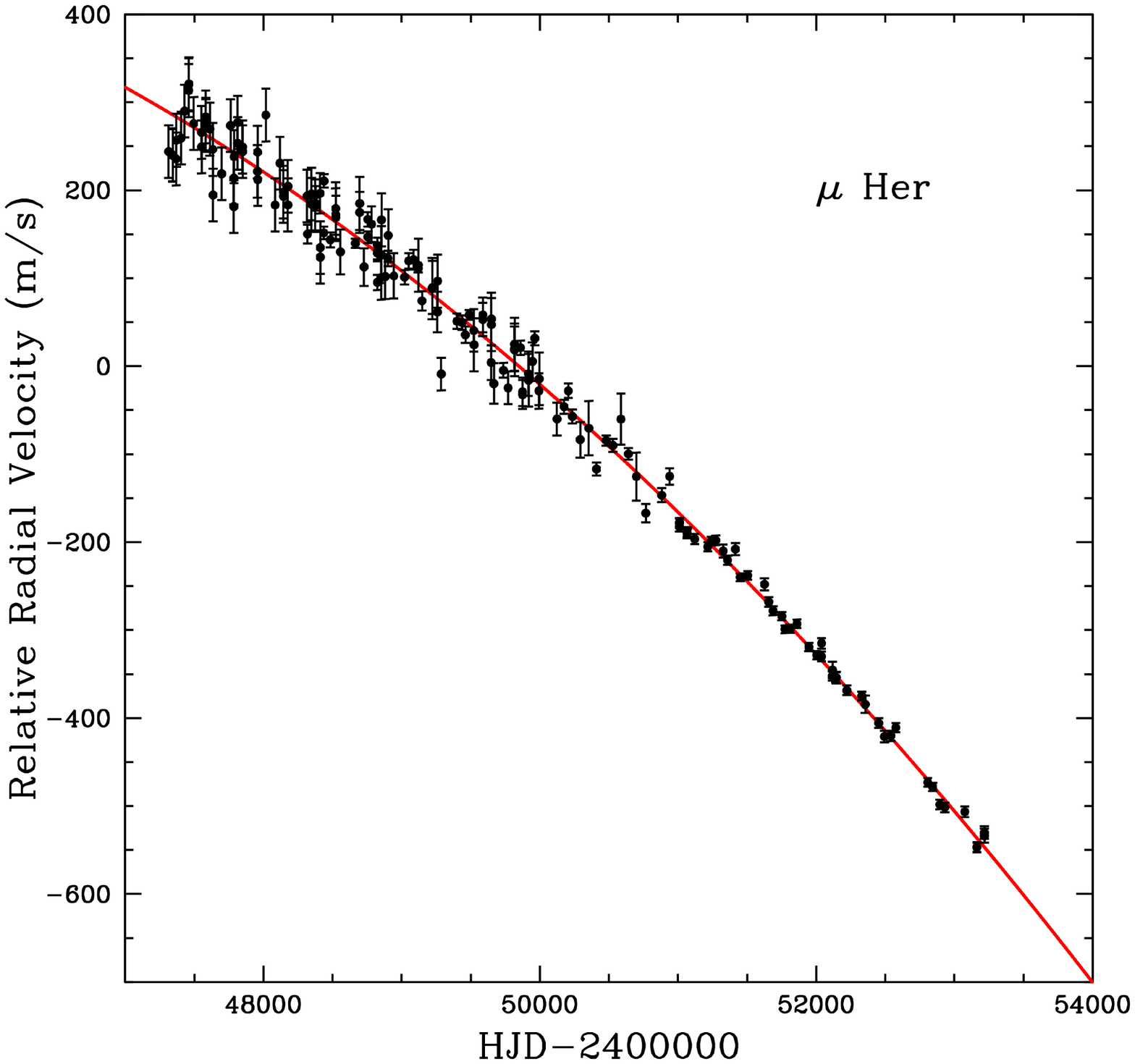}{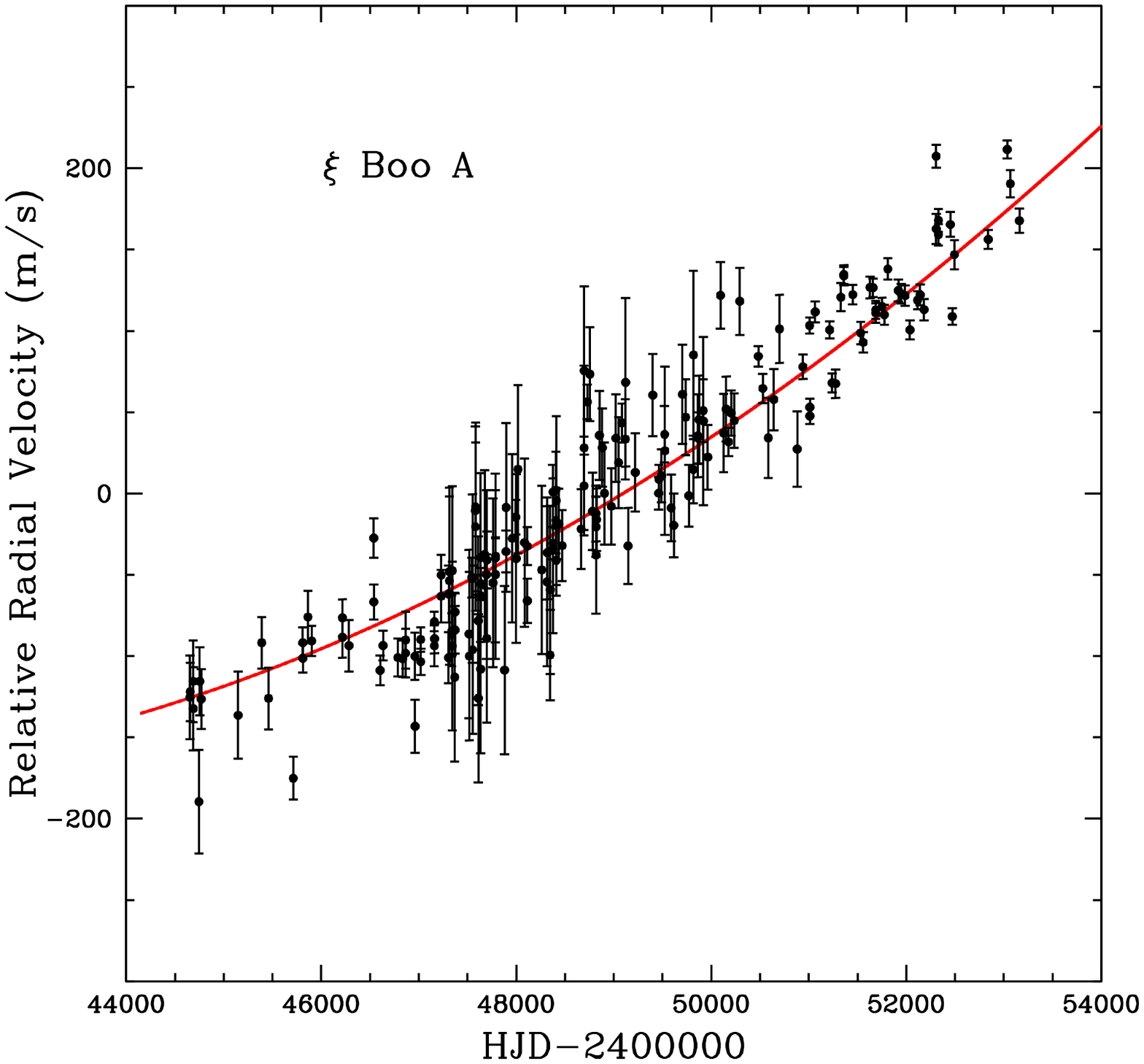}
\caption{Quadratic fits to $\mu$ Her data (left panel) and $\xi$ 
Boo A (right panel). }
\end{figure}

\clearpage
\begin{figure}
\plotone{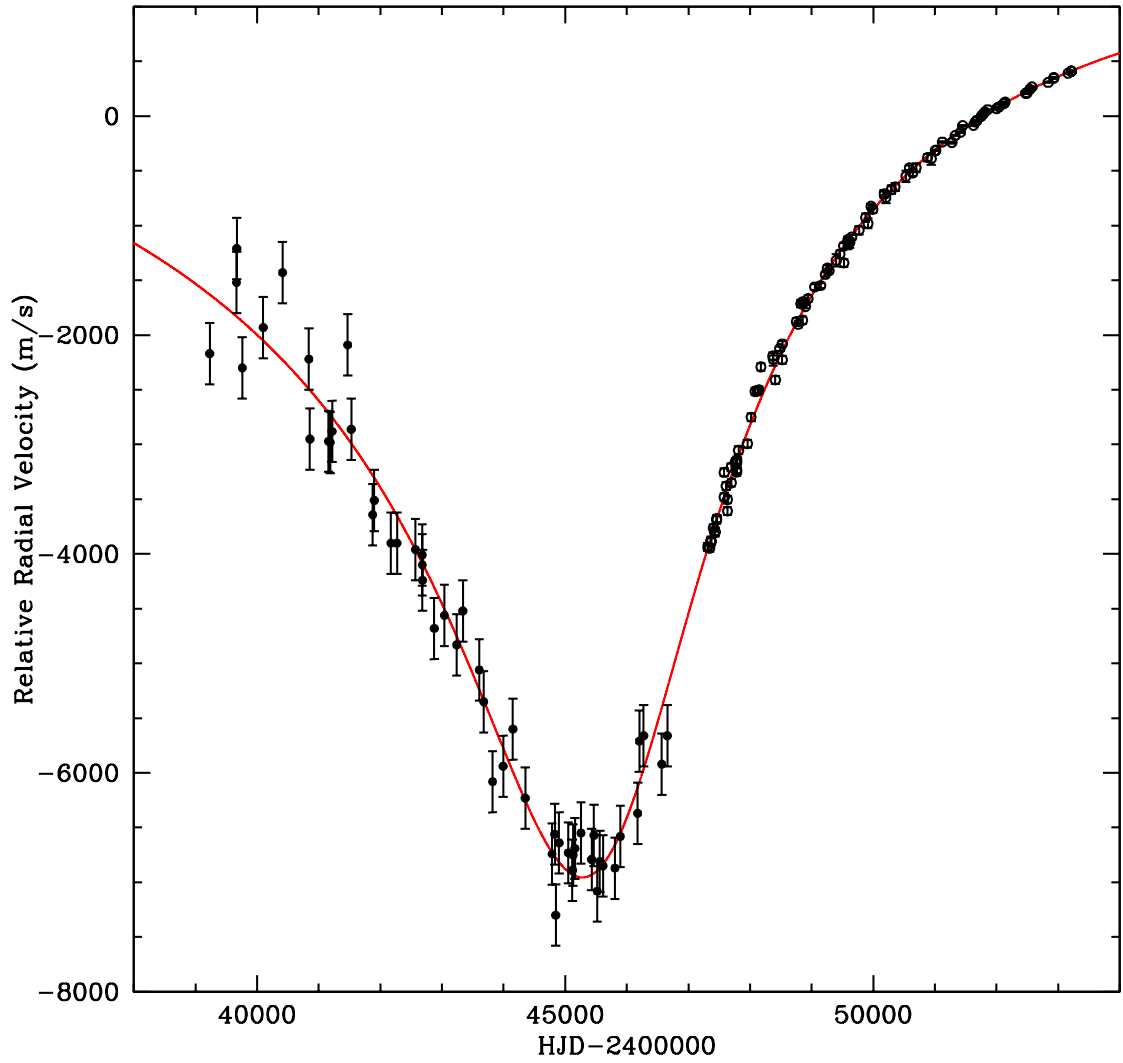}
\caption{Orbital solution for the visual binary 70 Oph A using data from 
\citet{batten84} (filled circles) and McDonald 
Observatory phase I-III (open circles).}
\end{figure}
\epsscale{1.00}

\clearpage
\begin{figure}
\plotone{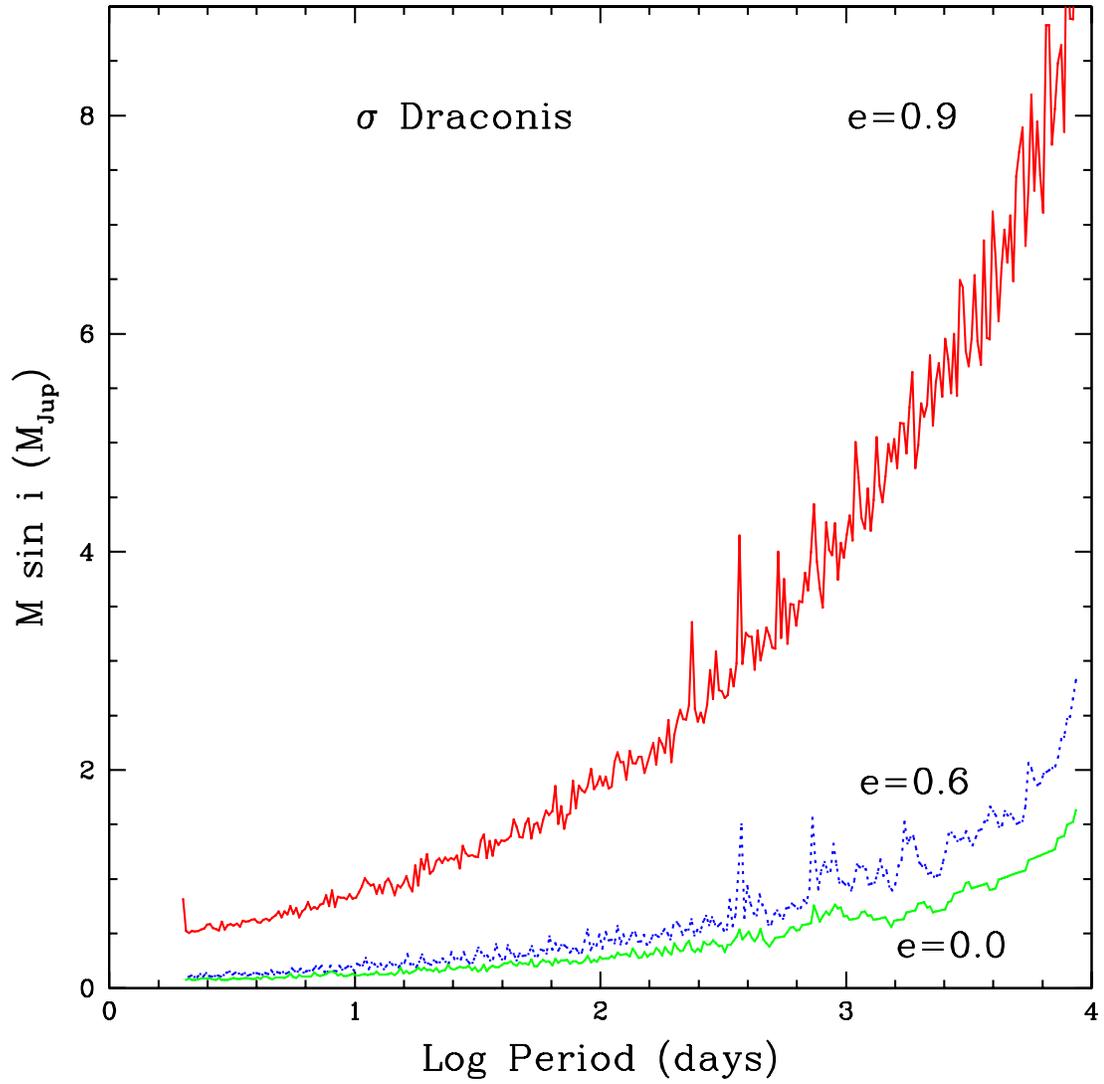}
\caption{The effect of eccentricity on limit determinations.  Allowing 
higher eccentricities reduces the sensitivity somewhat, due to the 
increased probability of unfortunately-phased observations. } 
\end{figure}

\begin{figure}
\plotone{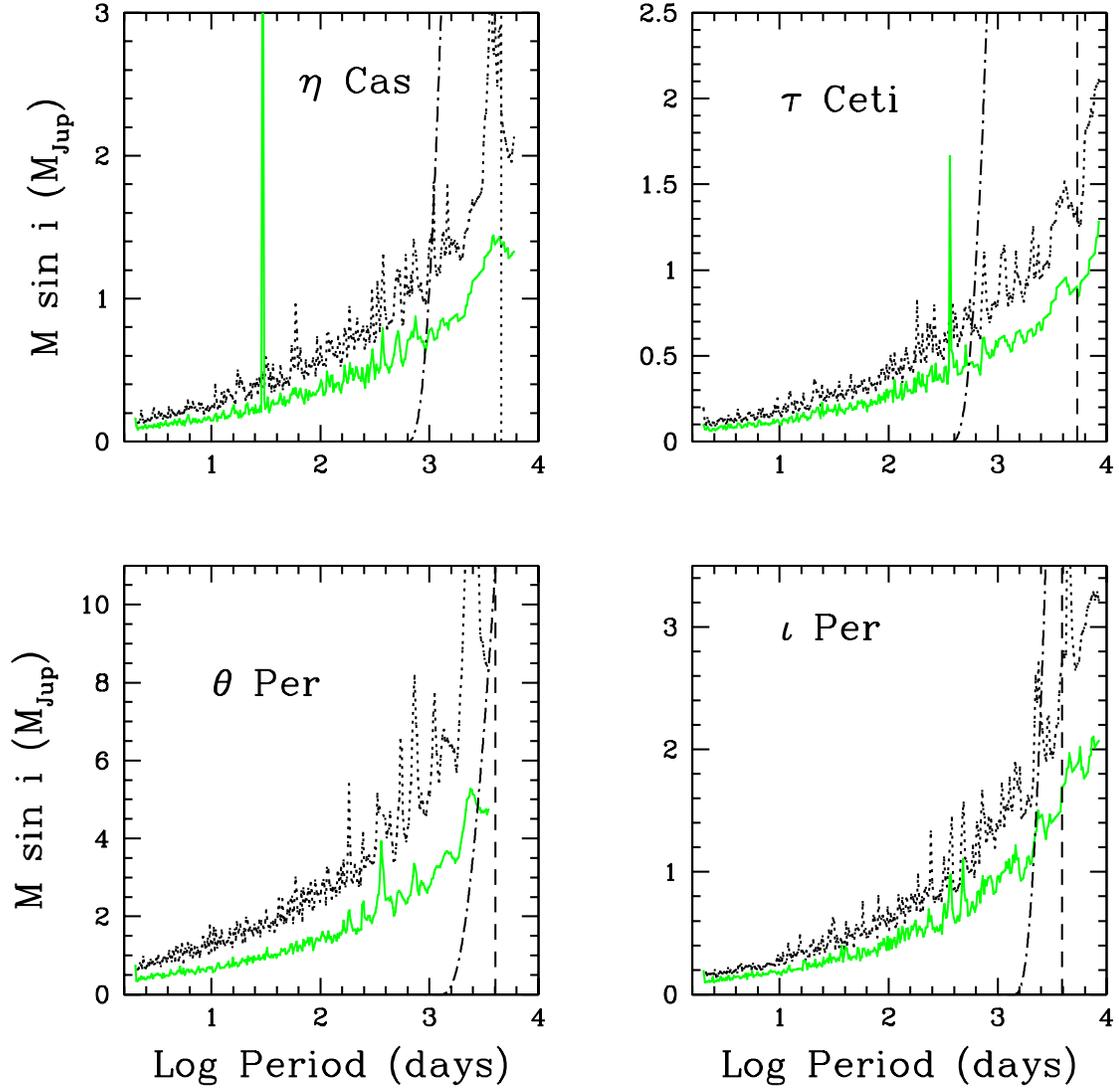}
\caption{Planetary companion limits for $\eta$ Cas, $\tau$ Cet, $\theta$
Per, and $\iota$ Per. The vertical dashed line indicates the 11.87 yr
orbital period of Jupiter. Planets in the parameter space above the
plotted points are excluded at the 99\% confidence level. The solid line
represents the companion limits for the zero-eccentricity case, and the
dotted line is for the case of $e=0.6$. The region left of the dot-dashed
line represents the set of perturbers at $e=0.2$ which would disrupt the
habitable zone. }
\end{figure}

\begin{figure}
\plotone{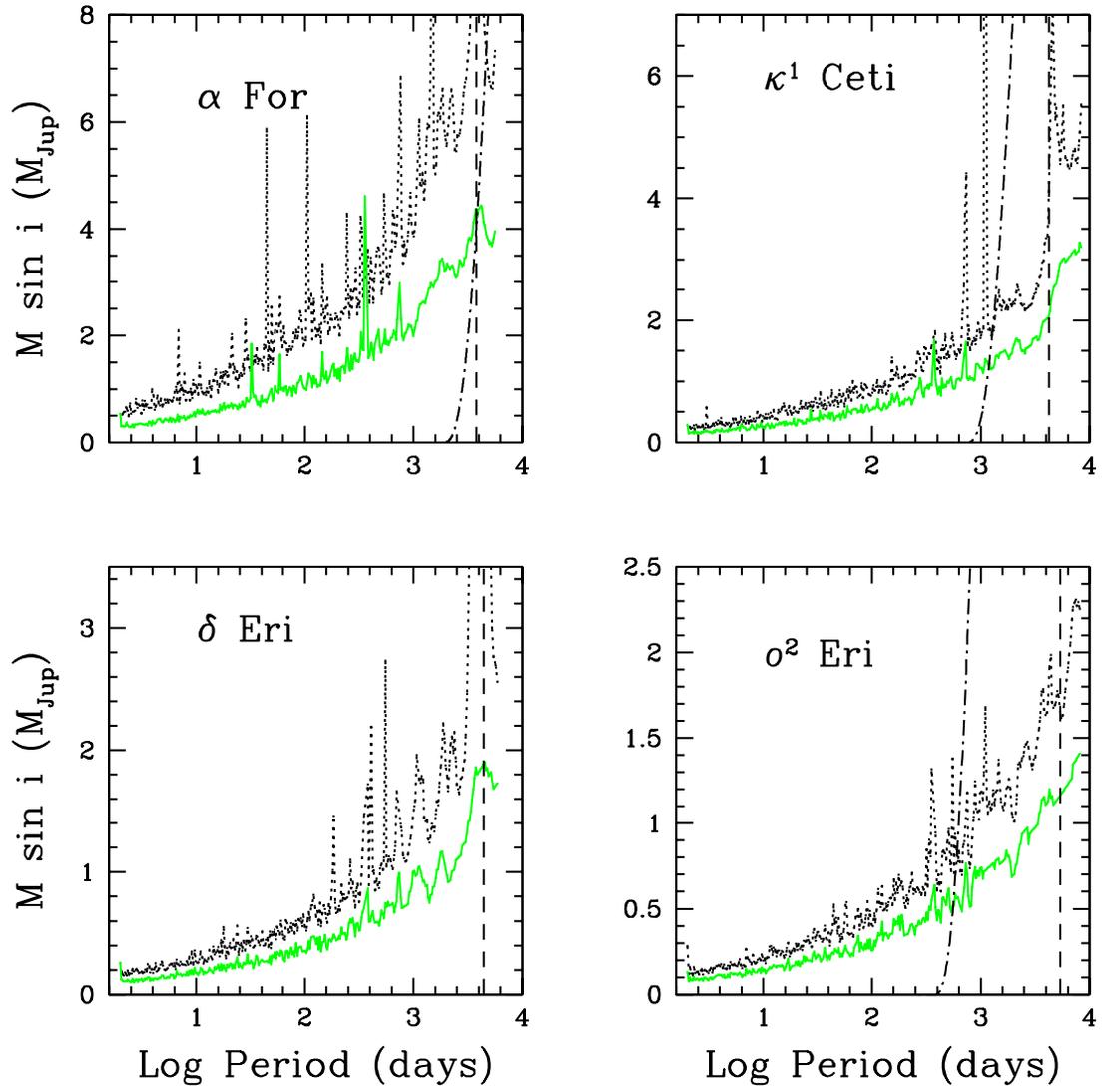}
\caption{Same as Figure 4, for $\alpha$ For, $\kappa^1$ Cet, 
$\delta$ Eri, and $o^2$ Eri. }
\end{figure}

\begin{figure}
\plotone{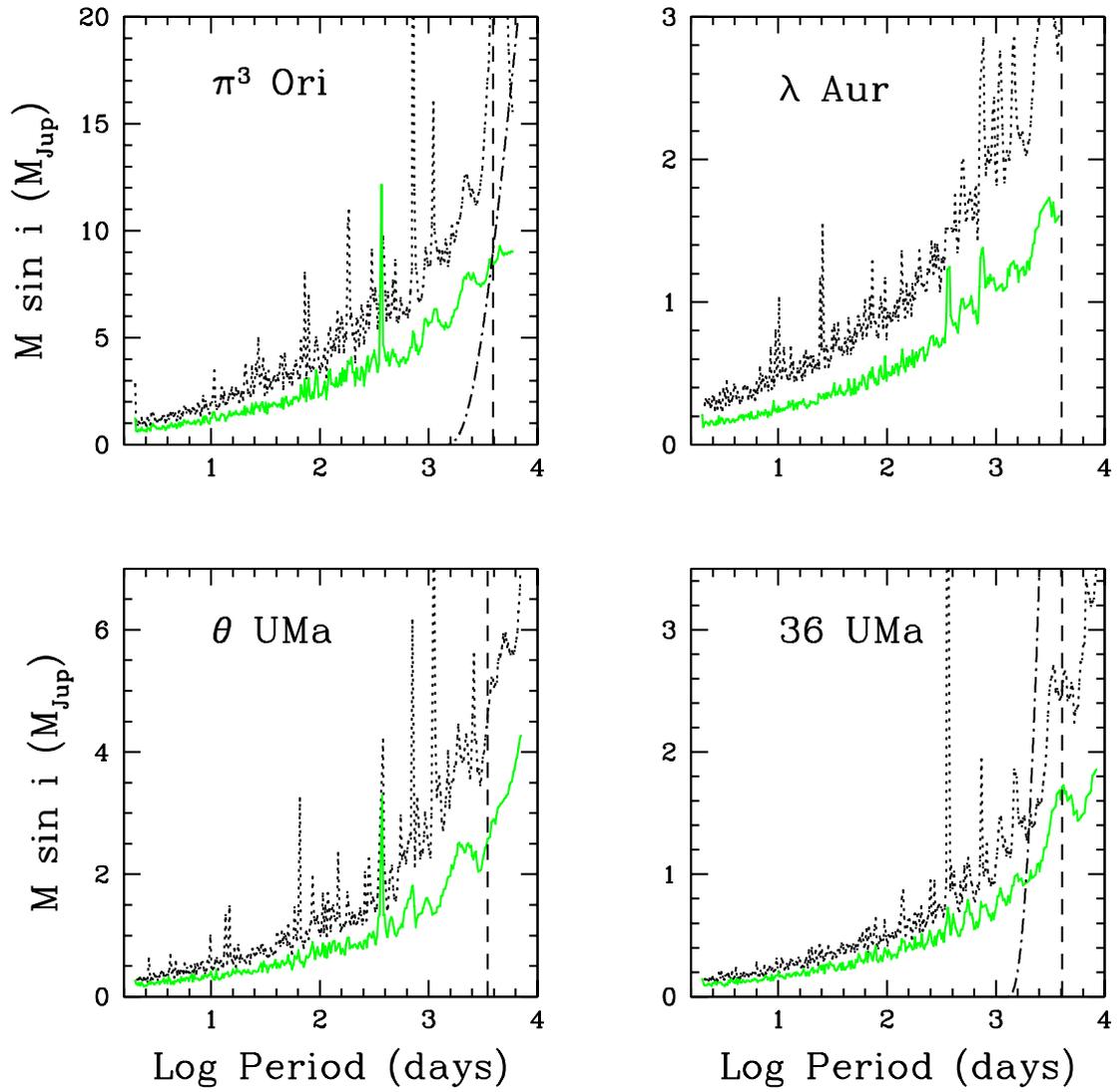}
\caption{Same as Figure 4, for $\pi^3$ Ori, $\lambda$ Aur, 
$\theta$ UMa, and 36 UMa. }
\end{figure}

\begin{figure}
\plotone{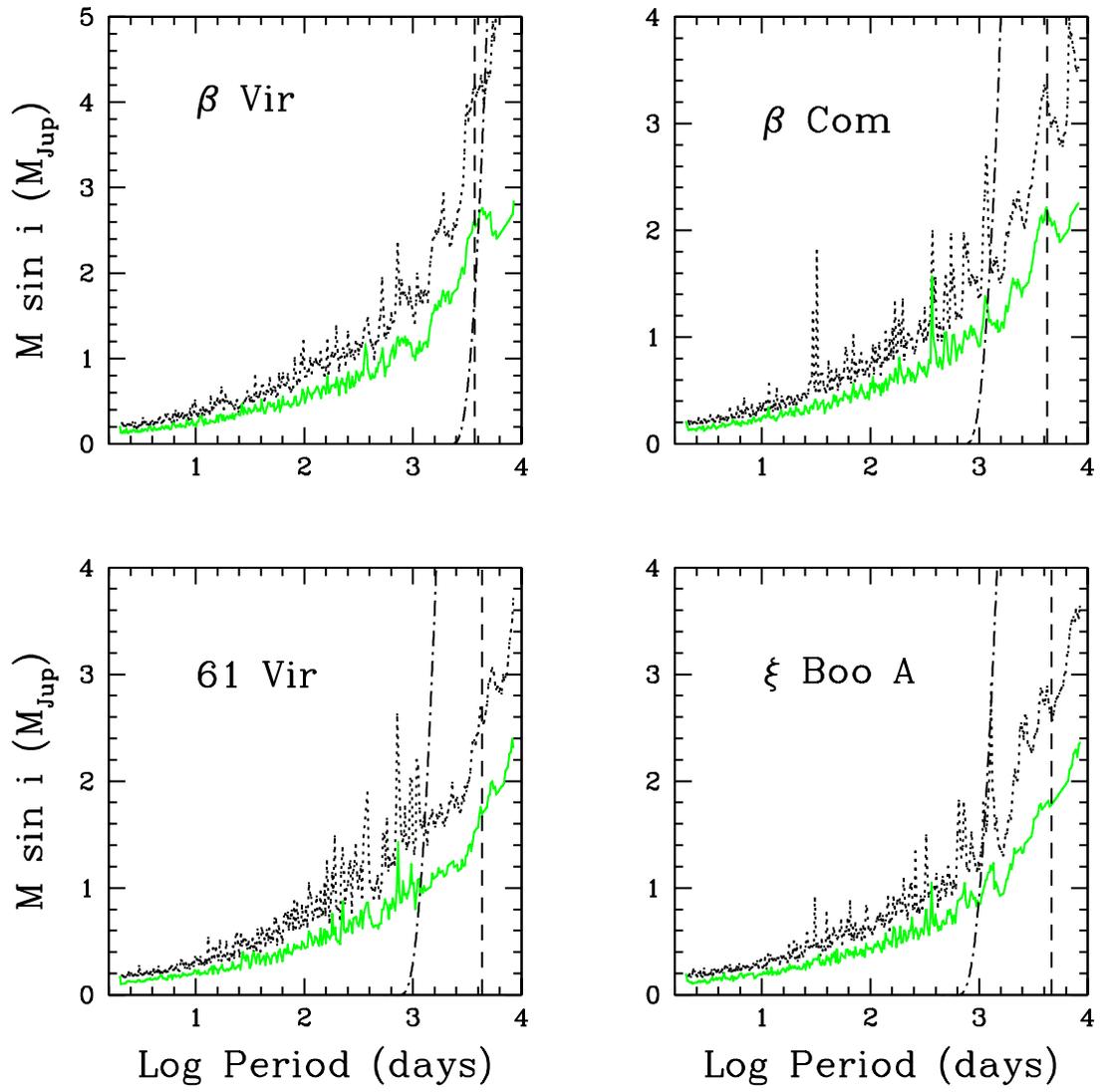}
\caption{Same as Figure 4, for $\beta$ Vir, $\beta$ Com, 61 Vir, 
and $\xi$ Boo A. }
\end{figure}

\begin{figure}
\plotone{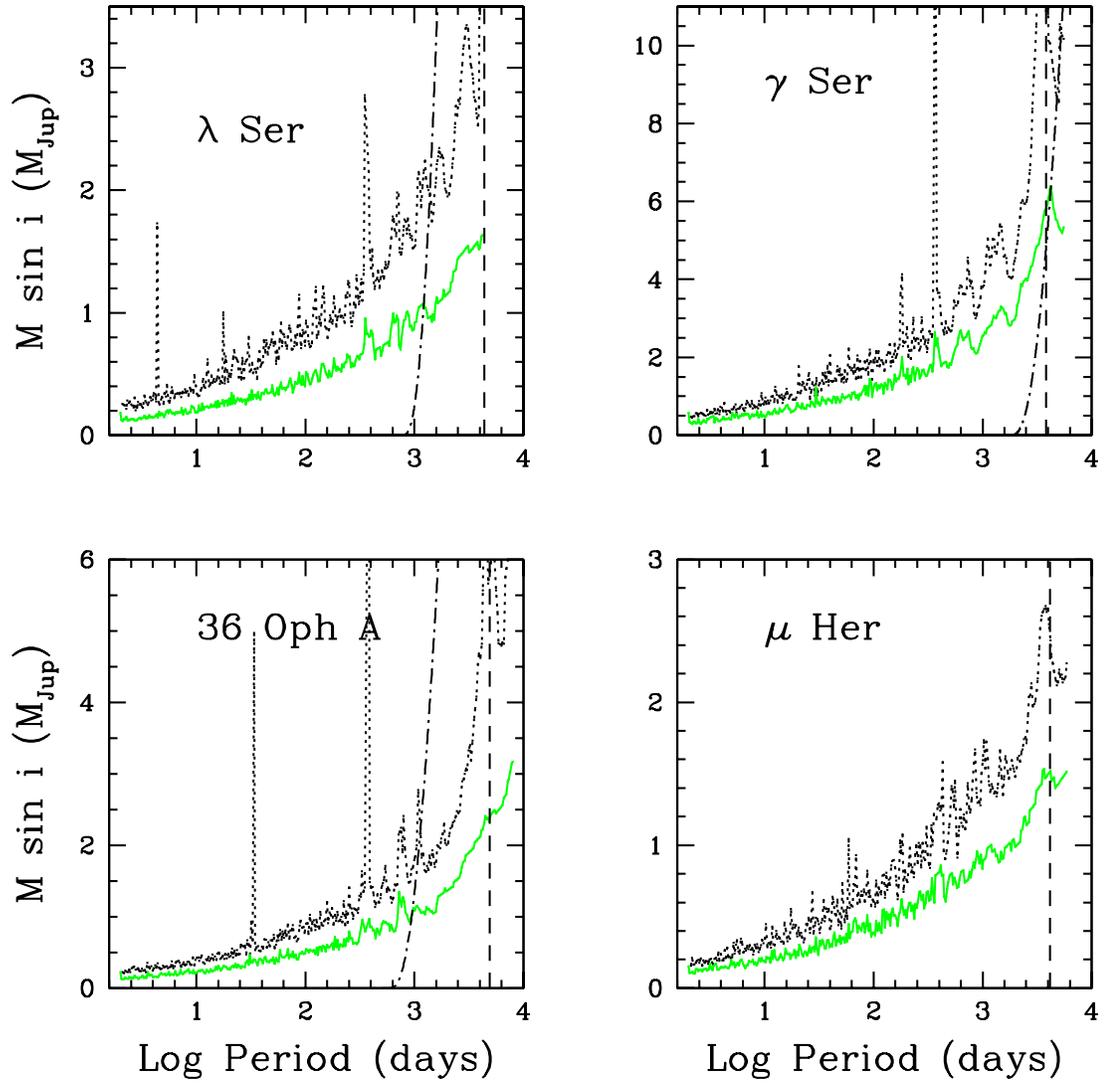}
\caption{Same as Figure 4, for $\lambda$ Ser, $\gamma$ Ser, 
36 Oph A, and $\mu$ Her. }
\end{figure}

\begin{figure}
\plotone{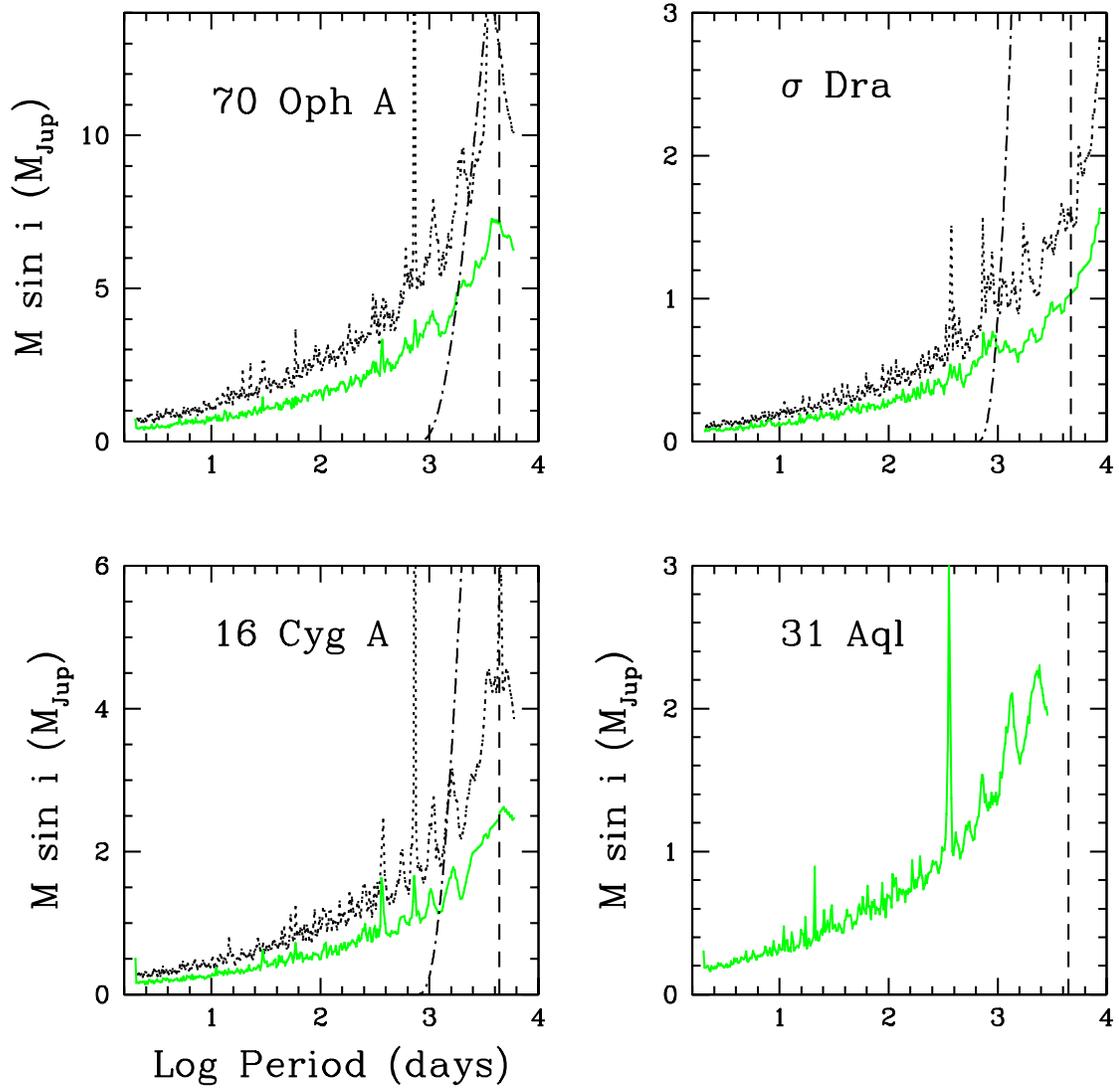}
\caption{Same as Figure 4, for 70 Oph A, $\sigma$ Dra, 16 Cyg A, 
and $\beta$ Aql. }
\end{figure}

\begin{figure}
\plotone{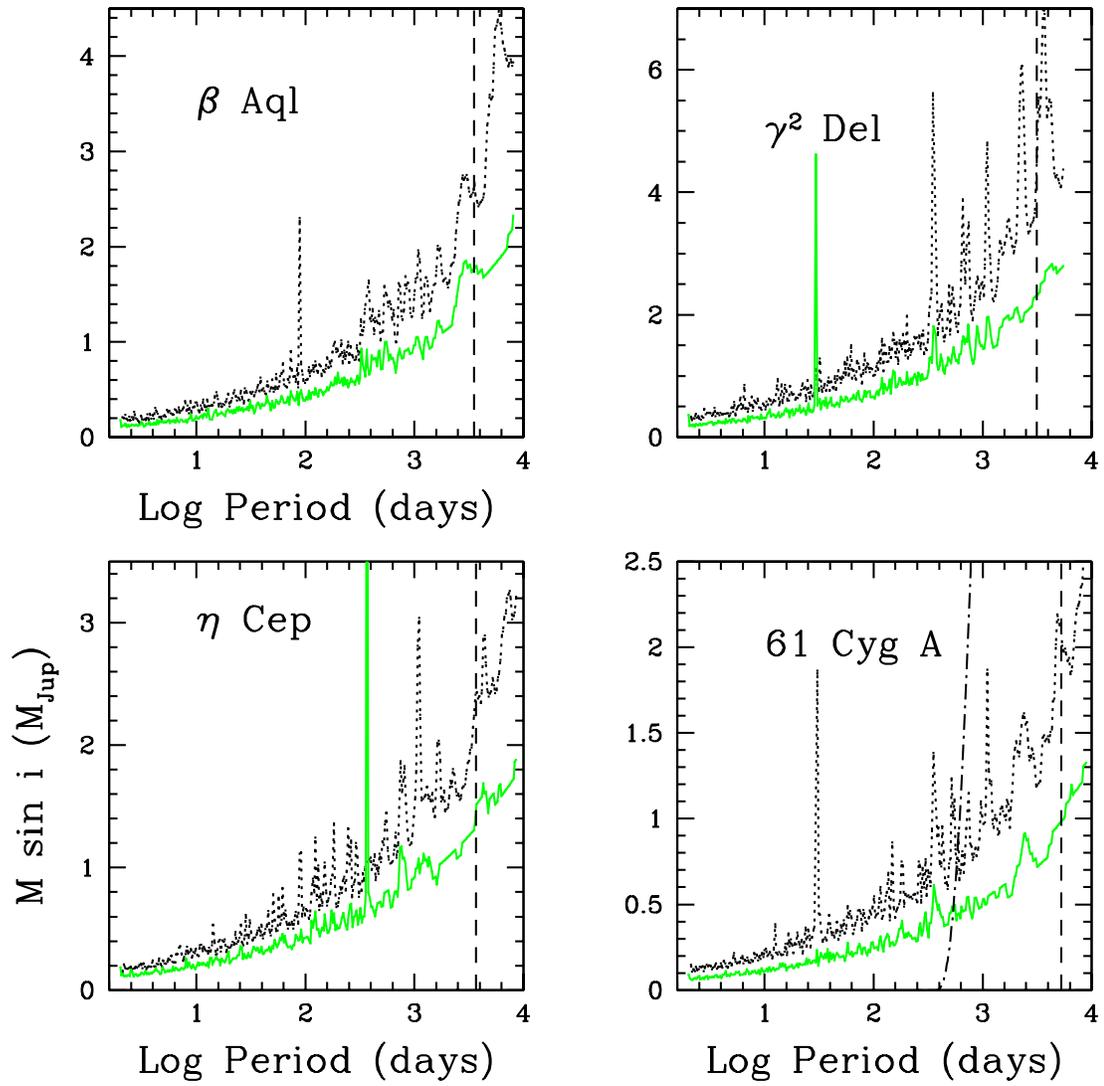}
\caption{Same as Figure 4, for 31 Aql, $\gamma^2$ Del, $\eta$
Cep, and 61 Cyg A. }
\end{figure}

\begin{figure}
\plotone{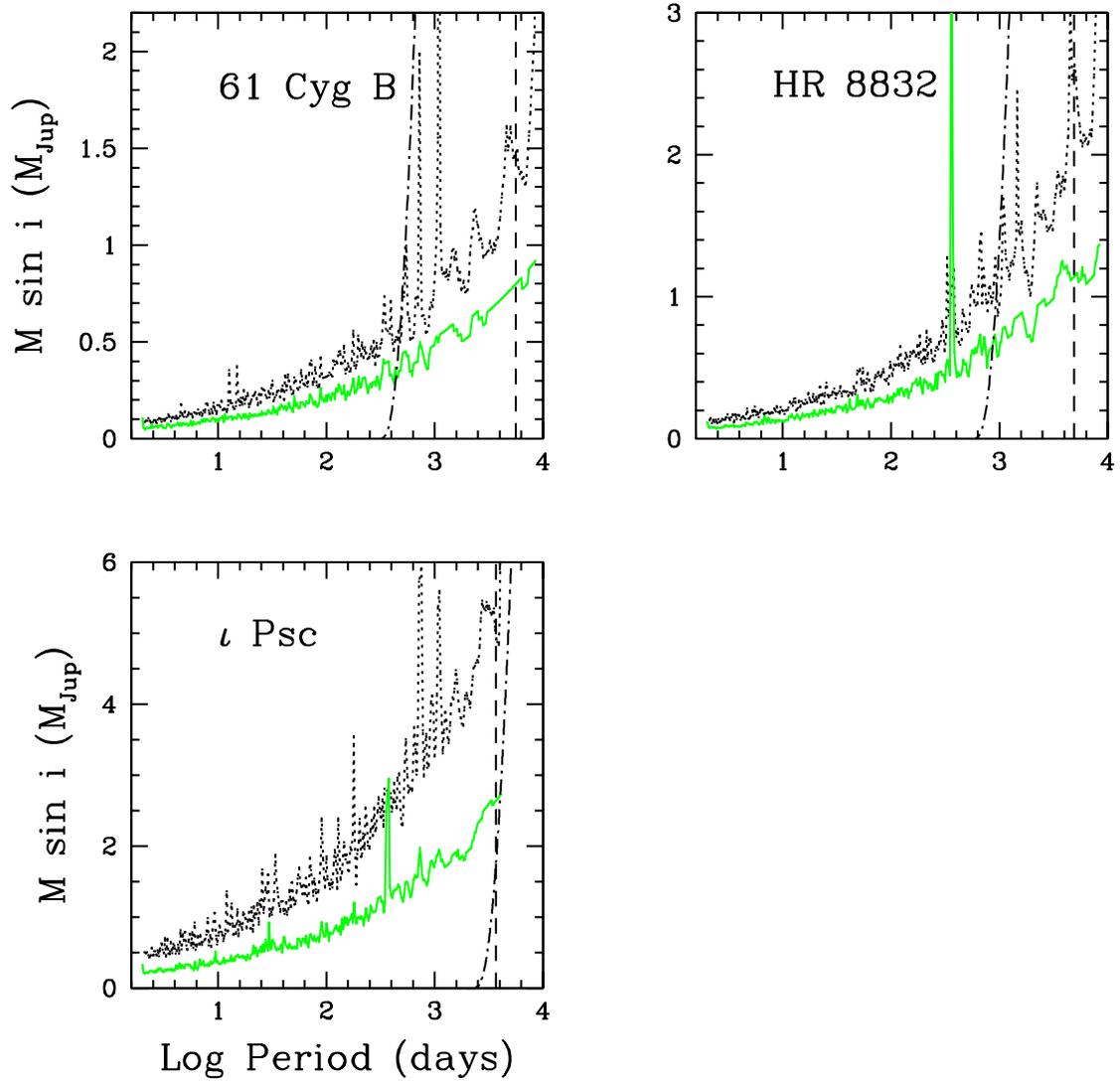}
\caption{Same as Figure 4, for 61 Cyg B, HR 8832, and $\iota$
Psc. }
\end{figure}

\begin{figure}
\plottwo{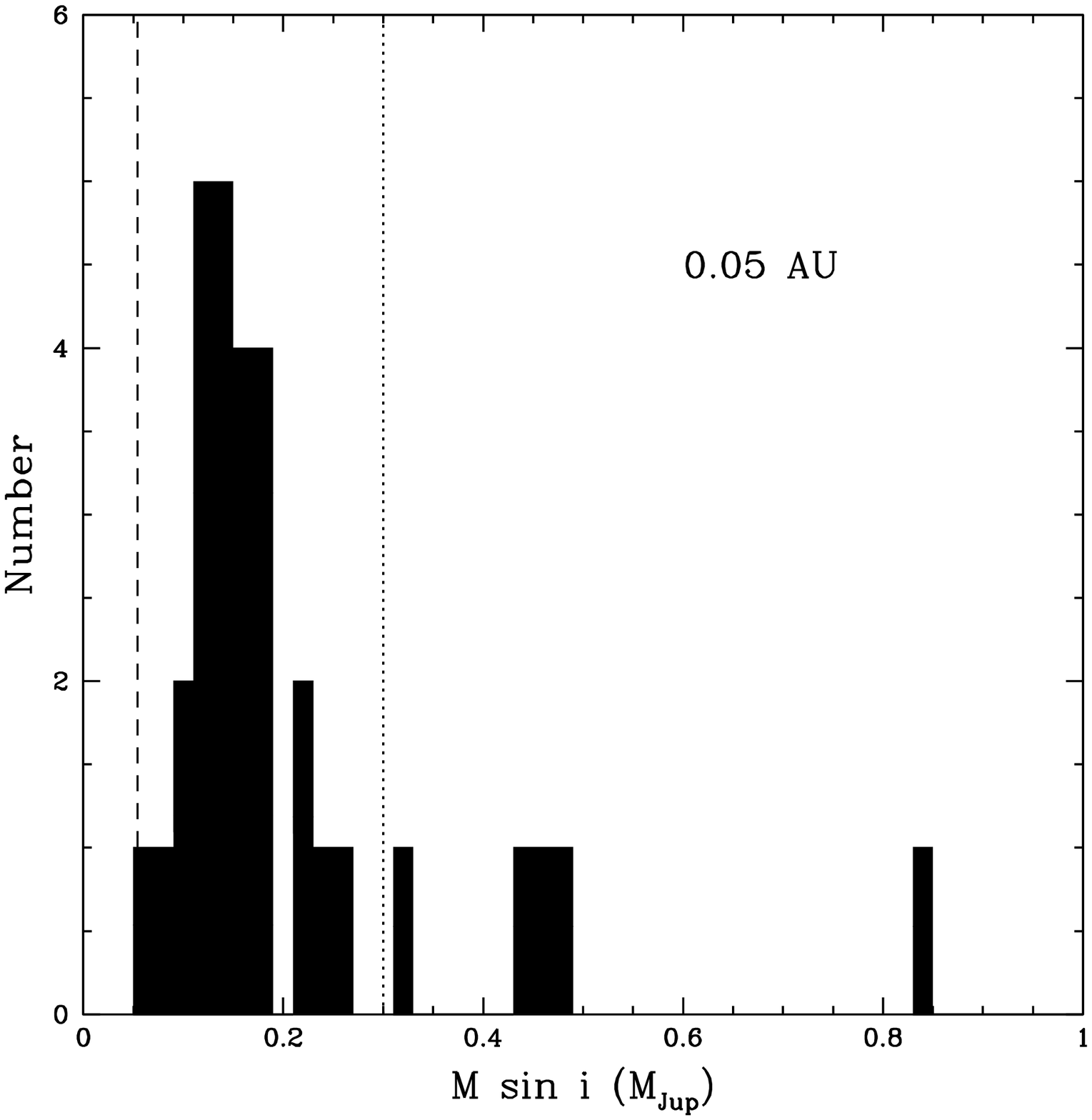}{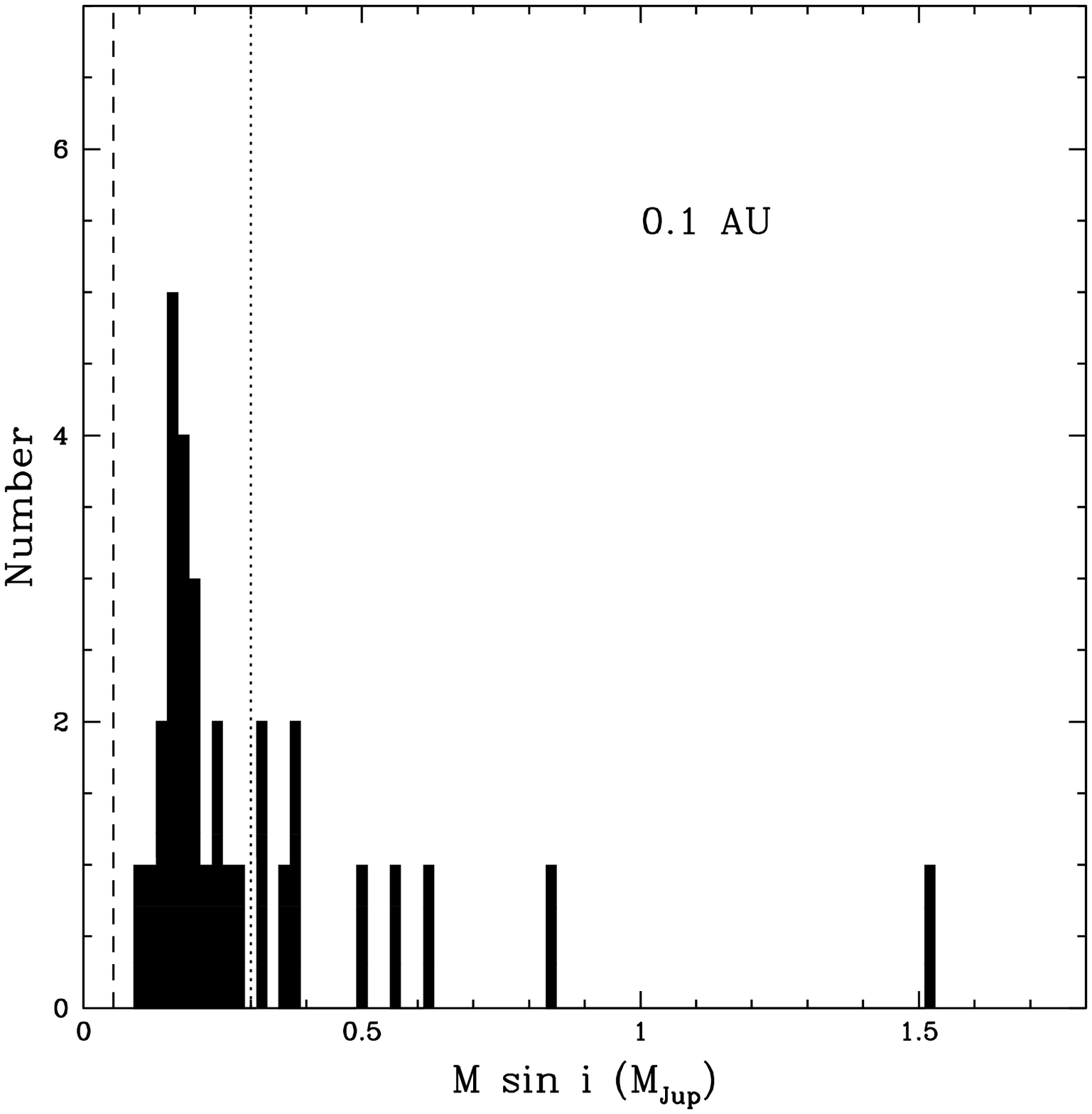}
\caption{Histogram showing the lowest M sin i of planets ruled out at
semimajor axes of 0.05 AU (left panel) and 0.1 AU (right panel).  The
dotted vertical line indicates the mass of Saturn, and the dashed line
indicates the mass of Neptune. }
\end{figure}

\begin{figure}
\epsscale{.8}
\plotone{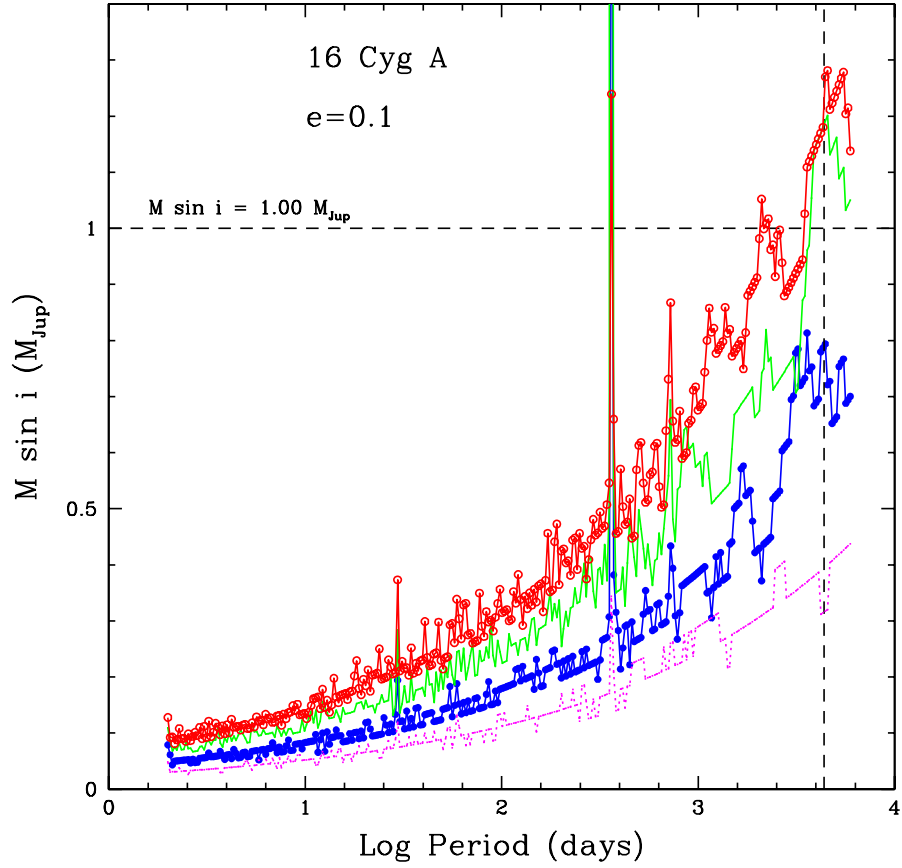}
\caption{Planetary companion limits for 16 Cyg A, using simulated data
matching our observation times, but with an rms of 14 m s$^{-1}$ (red,
open circles), 11 m s$^{-1}$ (green, solid line), 8 m s$^{-1}$ (blue,
filled circles) and 5 m s$^{-1}$ (magenta, dotted line).  The vertical
dashed line indicates the 11.87 yr orbital period of Jupiter. Planets in
the parameter space above the plotted points are excluded at the 99.9\%
confidence level.  This shows that a Jupiter analog could be ruled out at
the 99.9\% level with a sufficiently long baseline of data similar in
quality to the McDonald Observatory phase 3.  }
\end{figure}


\end{document}